\documentclass[
 twocolumn,
 preprintnumbers,
 amsmath,amssymb,
 aps,
 superscriptaddress,
]{revtex4-1}

\usepackage{times}
\usepackage{stmaryrd}
\usepackage[normalem]{ulem}
\usepackage{comment}
\usepackage[usenames,dvipsnames]{color}
\definecolor{dgreen}{rgb}{0.0, 0.5, 0.0}

\usepackage{color}
\usepackage[dvipdfmx]{graphicx}
\usepackage{bm}
\usepackage{here}
\usepackage{url}
\usepackage{subfigure}
\usepackage{mhchem}
\usepackage{comment}
\usepackage{siunitx}

\newcommand{\todayd}{\the\year/\the\month/\the\day}
\newcommand{\del}{\partial}

\newcommand{\bel}{\begin{easylist}}
\newcommand{\eel}{\end{easylist}}

\newcommand{\eref}[1]{Eq.~\eqref{#1}}

\newcommand{\fref}[1]{Fig.~\ref{#1}}

\newcommand{\ccite}[1]{Ref.~\cite{#1}}

\newcommand{\bkt}[1]{\left\langle#1\right\rangle}

\newcommand{\sumtwo}[2]%
{\mathop{\sum_{#1}}_{#2}}
\newcommand{\sumthree}[3]%
{\mathop{\mathop{\sum_{#1}}_{#2}}_{#3}}
\newcommand{\sumfour}[4]%
{\mathop{\mathop{\mathop{\sum_{#1}}_{#2}}_{#3}}_{#4}} 
\newcommand{\prodtwo}[2]%
{\mathop{\prod_{#1}}_{#2}}
\newcommand{\mintwo}[2]%
{\mathop{\min_{#1}}_{#2}}
\newcommand{\maxtwo}[2]%
{\mathop{\max_{#1}}_{#2}}
\newcommand{\maxthree}[3]%
{\mathop{\mathop{\max_{#1}}_{#2}}_{#3}}
\newcommand{\limtwo}[2]%
{\mathop{\lim_{#1}}_{#2}}
\newcommand{\suptwo}[2]%
{\mathop{\sup_{#1}}_{#2}}
\newcommand{\supthree}[3]%
{\mathop{\mathop{\sup_{#1}}_{#2}}_{#3}}
\newcommand{\supfour}[4]%
{\mathop{\mathop{\mathop{\sup_{#1}}_{#2}}_{#3}}_{#4}} 
\newcommand{\inftwo}[2]%
{\mathop{\inf_{#1}}_{#2}}
\newcommand{\infthree}[3]%
{\mathop{\mathop{\inf_{#1}}_{#2}}_{#3}}
\newcommand{\inffour}[4]%
{\mathop{\mathop{\mathop{\inf_{#1}}_{#2}}_{#3}}_{#4}} 

\newcommand\calD{{\mathcal D}}

\newcommand\calT{{\mathcal T}}

\newcommand{\bsd}{\boldsymbol{d}}
\newcommand{\bse}{\boldsymbol{e}}
\newcommand{\bsf}{\boldsymbol{f}}

\newcommand{\bsj}{\boldsymbol{j}}

\newcommand{\bsp}{\boldsymbol{p}}

\newcommand{\bsu}{\boldsymbol{u}}
\newcommand{\bsv}{\boldsymbol{v}}

\newcommand{\bsx}{\boldsymbol{x}}

\newcommand{\bsF}{\boldsymbol{F}}

\newcommand{\bsxi}{\boldsymbol{\xi}}

\newcommand{\bsna}{\boldsymbol{\nabla}}

\newcommand{\tot}{\mathrm{tot}}
\newcommand{\hk}{\mathrm{hk}}
\newcommand{\ex}{\mathrm{ex}}
\newcommand{\ssrm}{\mathrm{ss}}

\begin{document}
\title{Thermodynamic Uncertainty Relations for Steady-State Thermodynamics}
\author{Takuya Kamijima}
\affiliation{
 Department of Applied Physics, The University of Tokyo, 7-3-1 Hongo, Bunkyo-ku, Tokyo 113-8656, Japan}
\author{Sosuke Ito}
\affiliation{
 Department of Physics, The University of Tokyo, 7-3-1 Hongo, Bunkyo-ku, Tokyo 113-0033, Japan}
\author{Andreas Dechant}
\affiliation{
 Department of Physics No. 1, Graduate School of Science, Kyoto University, Kyoto 606-8502, Japan}
\author{Takahiro Sagawa}
\affiliation{
 Department of Applied Physics, The University of Tokyo, 7-3-1 Hongo, Bunkyo-ku, Tokyo 113-8656, Japan}
\affiliation{
 Quantum-Phase Electronics Center (QPEC), The University of Tokyo, 7-3-1 Hongo, Bunkyo-ku, Tokyo 113-8656, Japan}

\date{\today}

\begin{abstract}
{A system can be driven out of equilibrium by both time-dependent and nonconservative forces, which gives rise to a decomposition of the dissipation into two non-negative components, called the excess and housekeeping entropy productions. We derive thermodynamic uncertainty relations for the excess and housekeeping entropy. These can be used as tools to estimate the individual components, which are in general difficult to measure directly. We introduce a decomposition of an arbitrary current into excess and housekeeping parts, which provide lower bounds on the respective entropy production. Furthermore, we also provide a geometric interpretation of the decomposition, and show that the uncertainties of the two components are not independent, but rather have to obey a joint uncertainty relation, which also yields a tighter bound on the total entropy production. We apply our results to two examples that illustrate the physical interpretation of the components of the current and how to estimate the entropy production.}
\end{abstract}


\maketitle
\textit{Introduction.---}
There have been vast developments in experimental techniques for microscopic systems \cite{Seifert2012,Parrondo-Horowitz-Sagawa,Ciliberto2017}, which makes it possible to measure the thermodynamic quantities of microscopic systems such as heat and work. This has {enabled direct applications of an emerging field of thermodynamics}, called stochastic thermodynamics \cite{seifert2008stochastic,Seifert2012}, where thermal fluctuation plays an important role in the nonequilibrium processes.
{A recently discovered nonequilibrium relation}, the thermodynamic uncertainty relation (TUR) \cite{Barato-Seifert2015,Gingrich-Horowitz2016,Gingrich-Rotskoff-Horowitz2017,proesmans2017discrete,dechant2018multidimensional,Brandner-Hanazato-Saito2018,Li-Horowitz-Gingrich-Fakhri2019,koyuk2019operationally,koyuk2020thermodynamic,Liu-Gong-Ueda2020,Dechant-Sasa2020,Otsubo-Ito-Dechant-Sagawa2020,Otsubo-Manikandan-Sagawa-Krishnamurthy2020} {states} that, in the short-time limit \cite{Otsubo-Ito-Dechant-Sagawa2020,Shiraishi-Saito-Tasaki2016},
\begin{align}
    \label{TUR}
    \sigma^\tot&\geq\frac{\left(j_d\right)^2}{\calD_d},
\end{align}
where $\sigma^\tot$ is the entropy production (EP) rate of the total system and  $j_d, \calD_d$ are the average and variance of a generalized current, respectively (the precise definition will be given later).
Since ${\left(j_d\right)^2}/{\calD_d}$ can be regarded as the precision of the current, inequality \eqref{TUR} {is a} tradeoff relation between the dissipation and the precision of the current.
{Measuring} the EP rate requires full statistics and it is {usually} not directly accessible in experiments. The TUR enables us to estimate the EP rate from the measurable average and variance of a current without assuming any specific model about the dynamics.

The second law of thermodynamics dictates that the total EP is always nonnegative at the level of ensemble average: $\sigma^\tot \geq 0$. In the presence of nonconservative driving, on the other hand, the ordinary second law does not give a tight bound, as dissipation does not disappear in the steady state, which is out of equilibrium due to the driving.  To refine the second law for such genuinely nonequilibrium situations, the total EP can be decomposed into two non-negative components, housekeeping (adiabatic) EP rate $\sigma^\hk$ and excess (nonadiabatic) EP rate $\sigma^\ex$, that is, $\sigma^\tot=\sigma^\hk+\sigma^\ex$ \cite{Hatano-Sasa,Speck-Seifert2005,3DFT-Esposito2010,Three-faces-Master-equation,Three-faces-Fokker-Planck}.
Here, $\sigma^\hk$ quantifies the intrinsic dissipation due to the nonconservative force, whereas $\sigma^\ex$ quantifies the dissipation due to the time-dependence of the system state. While these components offer detailed information about the nonequilibrium process, it is often hard to measure them directly in experiments.

In this Letter, we derive a generalized TUR for the housekeeping and excess EP rates of overdamped dynamics, by introducing two generalized currents: the housekeeping and excess current.
Just as the usual current has information about the dissipation, the introduced currents are nonequilibrium quantities possessing information about the corresponding EP rates. The generalized TUR has a geometrical representation connecting the EP rates and the currents, which we refer to as the projective TUR.
As a corollary of the projective TUR, it leads to the two TURs corresponding to the housekeeping and excess parts (as also discussed in \ccite{Dechant-Sasa-Ito2022}), while indicating that the two TURs are not independent of each other. 

The projective TUR further gives a tighter bound on the total EP than the conventional TUR, \eref{TUR}. It turns out that for a particular choice of the current coefficient, the generalized currents reduce to the usual current, implying that the housekeeping and excess EP rates can be estimated only by using directly measurable quantities. We note that such treatment cannot be straightforwardly extended to Markov jump processes, and a further modification of the current variance is required.

Specifically, we demonstrate the application of our TUR to two paradigmatic examples of time-dependent systems driven by nonconservative forces. The first example is two-dimensional Brownian motion, where a harmonic potential exerts a conservative force in the radial direction and a nonconservative force is exerted in the circumferential direction. In this setup, the housekeeping (excess) current coincides with the physical current in the circumferential (radial) direction. The second example is the so-called rocking ratchet \cite{Bartussek1994,Reimann2002}, where the system is spatially periodic and a time-periodic force is driving a particle current. We estimate the excess entropy production {using the known expression of the instantaneous steady-state} probability distribution.

\textit{Main result.---}
We consider the general overdamped Langevin equation
\begin{align}
    \label{Langevin eq}
    \dot{\bsx}=\bsF(\bsx(t), t)+\sqrt{2}G(t)\bsxi(t),
\end{align}
where {$\bsF$} is the drift term and $\bsxi$ is mutually independent white Gaussian noise, {and its components satisfy $\bkt{\xi_i}=0\ (i=1,\cdots,d)$ and $\bkt{\xi_i(t)\xi_j(t')}=\delta_{ij}\delta(t-t')$. $G(t)$ represents the strength of the noise.} 
The corresponding Fokker-Planck equation is written as 
\begin{align}
    \label{Fokker-Planck eq}
    \partial_t p(\bsx, t)&=-\bsna^T\bsj(\bsx, t),\\
    \bsj(\bsx, t)&=(\bsF(\bsx, t)-D\bsna)p(\bsx, t),
\end{align}
{where $D(t):=G(t)G(t)^T$ is the diffusion matrix ($^T$ denotes the transpose of a vector or matrix). We assume that $G$ has full rank and does not depend on $\bsx$.}
In addition, when the system is coupled to multiple reservoirs, it is assumed that $D$ is diagonal, i.e., there is no direct interaction between the reservoirs.

{If the drift term $\bsF$ contains only conservative forces and the system is coupled to a single reservoir, the system will relax to its equilibrium state, which satisfies the detailed balance condition. On the other hand, if this condition is violated and the parameters of the dynamics are fixed at time $t$, the system will relax to its steady state determined by the parameters, called the instantaneous steady state. We denote as $p^\ssrm(x,t)$ the probability distribution of the instantaneous steady state, where the $t$-dependency represents the parameters at time $t$.}

Using the mean local velocity \cite{Seifert2012}, $\bsv(\bsx,t):=\bsj(\bsx, t)/p(\bsx, t)=\bsF(\bsx, t)-D\bsna \ln p(\bsx, t)$, we introduce the housekeeping and excess currents as $j_d^\hk:=\int d\bsx \bsd^T \bsv^\ssrm p,\ j_d^\ex:=\int d\bsx \bsd^T (\bsv - \bsv^\ssrm)p$, respectively, {where $\bsd(\bsx,t)$ and $\bsv^\ssrm(\bsx,t)$ denote the current coefficient and the mean local velocity of the instantaneous steady state.}
The housekeeping current comprises the deviation of the (instantaneous) steady state from the equilibrium state, and the excess current comprises the deviation of the nonequilibrium state from the steady state.
$j_d^\hk=j_d,\ j_d^\ex=0$ holds for the steady state, and $j_d^\ex=j_d,\ j_d^\hk=0$ holds for the system with the detailed balance condition.
By definition, the total current is decomposed into these currents as $j_d^\hk+j_d^\ex=j_d=\int d\bsx \bsd^T \bsv p$.

Our main result, the projective TUR, is now stated as
\begin{align}
    \label{projective TUR}
    \frac{\left(j_d^\hk\right)^2}{\sigma^\hk}
    +\frac{\left(j_d^\ex\right)^2}{\sigma^\ex}
    \leq \calD_d,
\end{align}
where $\calD_d:=\int d\bsx \bsd^T D \bsd$ is the (time-rescaled) variance of the current. 
As a corollary of the projective TUR \eqref{projective TUR}, we can deduce the housekeeping and excess TURs:
\begin{align}
    \label{respective TUR}
    \sigma^\hk\geq\frac{\left(j_d^\hk\right)^2}{\calD_d},\ 
    \sigma^\ex\geq\frac{\left(j_d^\ex\right)^2}{\calD_d},
\end{align}
which have the same form as the conventional TUR \eqref{TUR} and have been discussed in \ccite{Dechant-Sasa-Ito2022}.
Since a non-negative term is removed from the left-hand side of \eref{projective TUR}, these TURs in \eqref{respective TUR} are looser than \eref{projective TUR}.
Importantly, the projective TUR \eqref{projective TUR} indicates that the TUR bounds for the housekeeping and excess EP rates are not independent of each other, in contrast to the looser version \eqref{respective TUR}. In fact, if we write the housekeeping EP rate as $\sigma^\hk=s{\left(j_d^\hk\right)^2}/{\calD_d}$ with some constant $s \geq 1$, then the excess EP rate is bounded as $\sigma^\ex\geq\frac{s}{s-1}{\left(j_d^\ex\right)^2}/{\calD_d}$, which is tighter than the inequality in \eqref{respective TUR}. {This tradeoff between the two components of the EP rate} {is a direct consequence of} the projective TUR \eqref{projective TUR}.

Furthermore, the projective TUR yields a TUR for the total dissipation as 
\begin{align}
    \label{max TUR}
    \sigma^{\textrm{tot}}&\geq\frac{1}{\calD_d}
    \max\{\left(j_d^\hk+j_d^\ex\right)^2,
    \left(j_d^\hk-j_d^\ex\right)^2\}.
\end{align}
When the product of the current components $j_d^\hk j_d^\ex$ is positive, this inequality simply reduces to the original TUR, \eref{TUR}. By contrast, when the product is negative, the TUR \eqref{max TUR} offers a tighter lower bound than \eref{TUR}. Note that \eref{max TUR} is always tighter than the bound obtained by simply summing up \eref{respective TUR}.

For Markov jump processes, inequalities \eqref{projective TUR} and \eqref{max TUR} are not valid, because the mean local velocity does not satisfy an orthogonality condition mentioned below (\eref{orthogonal condition}). Instead, a counterpart of \eref{respective TUR} holds for Markov jump processes, requiring a modification of the variance of the current (see Sec. II of Supplemental Material).

\textit{Derivation.---}
\begin{figure}
    \centering
    \includegraphics[width=0.5\linewidth]{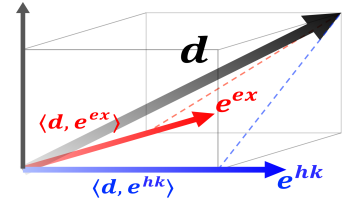}
    \caption{\label{fig: projection}
    Sketch of the projections of the current coefficient $\bsd$. Each axis represents the element of the basis of the vector field. The housekeeping (excess) current is represented by the blue (red) line segment, that is, the projected component to the housekeeping (excess) vector field. The length of the coefficient coincides with the square root of the current variance.
    }
\end{figure}
We derive inequality \eqref{projective TUR} by projecting the current coefficient $\bsd$ into the housekeeping
and excess vector fields.
For vector fields $\bsu(\bsx),\bsu'(\bsx)$, we define an inner product as $\bkt{\bsu,\bsu'}:=\int d\bsx \bsu^T D\bsu' p$ and the norm as $||\bsu||:=\sqrt{\bkt{\bsu,\bsu}}$. At each time, the orthogonality condition
\begin{align}
    \label{orthogonal condition}
    \bkt{D^{-1}\bsv^\ssrm, D^{-1}(\bsv-\bsv^\ssrm)}=0
\end{align}
holds \cite{Three-faces-Fokker-Planck}, where {$D^{-1}$ denotes the inverse matrix of $D$}. That is, the housekeeping and excess thermodynamic forces, $D^{-1}\bsv^\ssrm$ and $D^{-1}(\bsv-\bsv^\ssrm)$, are orthogonal to each other in terms of the inner product $\bkt{\cdot,\cdot}$.
Using this relation, each EP rate can be written as $\sigma^{\textrm{tot}}=||D^{-1}\bsv||^2,\ \sigma^\hk=||D^{-1}\bsv^\ssrm||^2,\ \sigma^\ex=||D^{-1}(\bsv-\bsv^\ssrm)||^2$.
For a time-integrated generalized current $J_d(t):=\int^t ds \bsd(\bsx(s), s)\circ\dot{\bsx}$ with $\circ$ denoting the Stratonovich product, the average and variance of the (instantaneous) current are given by $j_d(t):=d\bkt{J_d(t)}/dt=\bkt{\bsd,D^{-1}\bsv}$ and $\calD_d(t):=\lim_{\tau\rightarrow0}{\textrm{Var}[J(t+\tau)-J(\tau)]}/{2\tau}=\bkt{\bsd,\bsd}$,
respectively \cite{Otsubo-Ito-Dechant-Sagawa2020}.

Then, we define the housekeeping and excess vector fields, $\bse^\hk$ and $\bse^\ex$, as the normalized corresponding thermodynamic forces, that is, $\bse^\hk(\bsx, t):={D^{-1}\bsv^\ssrm}/{||D^{-1}\bsv^\ssrm||},\ \bse^\ex(\bsx, t):={D^{-1}(\bsv-\bsv^\ssrm)}/{||D^{-1}(\bsv-\bsv^\ssrm)||}$.
These vector fields satisfy $\bkt{\bse^\hk,\bse^\hk}=\bkt{\bse^\ex,\bse^\ex}=1$ and $\bkt{\bse^\hk,\bse^\ex}=0$ due to the normalization and the orthogonality condition (\eref{orthogonal condition}).
Consequently, the measurable condition is derived for the current components:
\begin{align}
    \label{hk coeff}
    \bsd\perp\bsv- \bsv^\ssrm\textrm{ or }
    \bsd\propto\bse^\hk
    \Longrightarrow j_d^\hk=j_d,\ j_d^\ex=0,\\
    \label{ex coeff}
    \bsd\perp\bsv^\ssrm\textrm{ or }
    \bsd\propto\bse^\ex
    \Longrightarrow j_d^\ex=j_d,\ j_d^\hk=0.
\end{align}
For vector fields $\bsu$ and $\bsu'$, we define orthogonality $\perp$ and proportionality $\propto$ as $\bsu\perp\bsu'\Longleftrightarrow\forall\bsx,\ \bsu(\bsx)^T\bsu'(\bsx)=0$ and $\bsu\propto\bsu'\Longleftrightarrow\forall\bsx,\ \bsu'(\bsx)=c\bsu(\bsx)$ ($c(\neq0)$ is a constant), respectively.
When the current coefficient $\bsd$ satisfies \eref{hk coeff} (\eref{ex coeff}), the housekeeping (excess) current reduces to the measurable current. 
In the illustration of our inequalities, the current coefficient is chosen to be measurable, taking advantage of the orthogonality $\bsv^\ssrm\perp\bsv-\bsv^\ssrm$ in the first example and easier computability of the steady-state components in the second example.

If we choose $\{\bse^\alpha\}=\{\bse^\hk,\bse^\ex,\cdots\}$ as an orthonormal basis of the space of the vector field,
the current coefficient $\bsd$ can be expanded as $\bsd=\sum_\alpha\bkt{\bsd,\bse^\alpha}\bse^\alpha$. Then, the projective inequality
\begin{align}
    \label{derivation of projective TUR}
    \bkt{\bsd,\bsd}=\sum_\alpha\bkt{\bsd,\bse^\alpha}^2\geq
    \bkt{\bsd,\bse^\hk}^2+\bkt{\bsd,\bse^\ex}^2
\end{align}
leads to our main inequality \eref{projective TUR} (see \fref{fig: projection}). As clear from this derivation, more elements of the basis tighten the projective TUR (see Sec. I of Supplemental Material), but the physical meaning of the corresponding current component is not apparent. The equality of \eref{projective TUR} is achieved if and only if the current coefficient $\bsd$ only has a housekeeping and/or excess component. In addition, if $\bsd$ has only one of these components, i.e., $\bsd\propto\bse^\hk\textrm{ or }\bse^\ex$, the equality of the corresponding TUR \eref{respective TUR} is achieved.

{Meanwhile, by multiplying \eref{projective TUR} and $\sigma^\tot=\sigma^\hk+\sigma^\ex$, we can derive \eref{max TUR} as
\begin{align}
    \sigma^{\textrm{tot}}\calD_d
    &=(j_d^\hk)^2+(j_d^\ex)^2
    +\frac{\sigma^\ex}{\sigma^\hk}(j_d^\hk)^2+\frac{\sigma^\hk}{\sigma^\ex}(j_d^\ex)^2\nonumber\\
    &\geq (j_d^\hk)^2+(j_d^\ex)^2
    +2\sqrt{(j_d^\hk)^2(j_d^\ex)^2},
\end{align}
where we apply the arithmetic-geometric mean inequality in the last line.}
Inequality \eqref{max TUR} can also be derived by using the orthogonality condition and the Cauchy-Schwaltz inequality (see Sec. I of Supplemental Material). The equality of \eref{max TUR} holds if and only if $\bsd$ only has a housekeeping and/or excess component and the (absolute) ratios of the EP and currents agree with each other, $\sigma^\ex/\sigma^\hk=|j_d^\ex/j_d^\hk|$. 

\begin{figure}[t]
    \centering
    \includegraphics[width=1.0\linewidth]{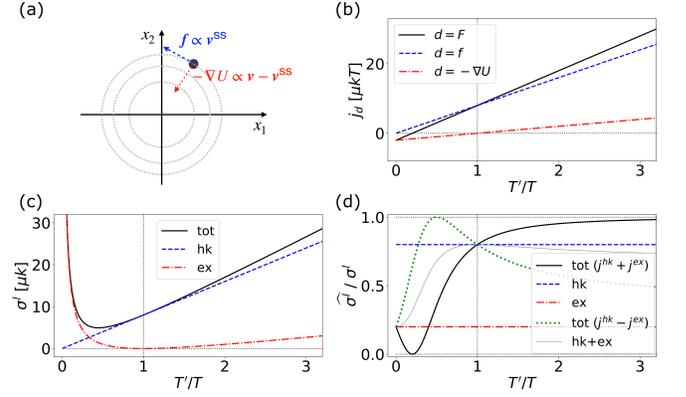}
    \caption{\label{fig: 2dflow} (a) A sketch of the nonconservative force in the circumferential direction and the conservative force in the radial direction acting on the Brownian particle. 
    (b) The average power injected into the system. The total current averages are plotted for the work coefficients. 
    (c) The individual EP rates. $T'$ represents the effective temperature of the system. $\sigma^\tot=\sigma^\hk$ holds at $T=T'$ because the system is in its steady state.  (d) The estimation of the EP rates using the TURs. $\hat{\sigma^l}\ (l=\tot,\hk,\ex)$ denotes the estimation of the corresponding EP. The blue and red lines represent the estimation based on \eref{respective TUR}. The gray line is the sum of these estimations, i.e., $\hat{\sigma^\tot}=\hat{\sigma^\hk}+\hat{\sigma^\ex}$. The black and green lines represent the estimations based on \eref{max TUR}. The parameters used in the calculation are $\mu=1.0, k=1.0, \kappa=2.0, T=1.5$.
    }
\end{figure}

\textit{Application to two-dimensional Brownian motion.---}
As an illustration of \eref{projective TUR}, we consider the two-dimensional Brownian motion sketched in \fref{fig: 2dflow}(a). The drift term and the noise strength are given by {$\bsF=-\bsna U+\bsf\ (\mu=1)$} and $G=\sqrt{T}I$ ($I$ is the identity matrix), respectively.
The Brownian particle moves under an isotropic harmonic potential $U(\bsx)=k({x_1}^2+{x_2}^2)/2\ (k>0)$, and there is a nonconservative force $\bsf(\bsx)=\kappa[-x_2, x_1]^T\ (\kappa>0)$ in the circumferential direction. This model has also been examined in \ccite{tomita1974, Nemoto-Sasa}. The steady-state distribution agrees with the canonical one {$\bsp^\ssrm={k}e^{-U/T}/{2\pi T}$}. Since $\bsj^\ssrm=\bsf p^\ssrm,\ \bsv^\ssrm=\bsf$, the mean local velocity coincides with the nonconservative force in the steady state.

Although the projective TUR holds for arbitrary states, we adopt the Gaussian distribution $p^{\textrm{G}}={k}e^{-U/T}/{2\pi T'}$ as the state of the system, where $T'$ can be regarded as the effective temperature of the system.
{In practice, such a state is obtained for a quench in either the trapping frequency or the temperature.}
Then, $\bsv=(1-T/T')(-\bsna U)+\bsf$ and $\bsv-\bsv^\ssrm=(1-T/T')(-\bsna U)$ hold,
and therefore $\bsv-\bsv^\ssrm$ is proportional to the conservative force.

We consider the work current per time (i.e., power) as the generalized current. For $\bsd=-\bsna U, \bsf, \bsF(=\bsf-\bsna U)$, the total currents correspond to the unit-time work exerted by the conservative force, the nonconservative force, and the total force, respectively. 
The average currents for individual coefficients are plotted in \fref{fig: 2dflow}(b). All of them are linear in terms of $T'$ and monotonically increasing.
Since $\bsf\propto\bse^\hk$ and $-\bsna U\propto\bse^\ex$, (or simply $-\bsna U\perp\bsf$), the housekeeping and excess currents for $\bsd=\bsF$ reduce to the total current for $\bsd=\bsf$ and $\bsd=-\bsna U$ respectively, i.e., {$j_{\bsd=\bsF}^\hk=j_{\bsd=\bsf}$ and $j_{\bsd=\bsF}^\ex=j_{\bsd=-\bsna U}$}. Thus, the black line in this figure agrees with the sum of the blue and red line because of the coefficient decomposition $\bsF = \bsf -\bsna U$ and the current decomposition $j_d = j_d^\hk+j_d^\ex$ for $\bsd=\bsF$. 

For the nonconservative force $\bsd=\bsf$, the average current is always positive (i.e., $j_d=j_d^\hk>0$) for $T'>0$, which shows that the work done by this force is always positive. 
On the other hand, for the conservative force $\bsd=-\bsna U$, a positive work is applied to the system for $T'>T$, and a negative work is applied to the system for $T'<T$. This means that when $T'>T$, for example, the distribution $p^{\textrm{G}}$ has higher (potential) energy than the steady state $p^\ssrm$ ($T'=T$), and the system dissipates the excess energy to the reservoir as the heat.


Finally, we estimate the EP rate (see \fref{fig: 2dflow} (c)) by applying \eref{projective TUR} to this model. The precision of the estimations $\widehat{\sigma^l}\ (l=\tot,\hk,\ex)$ with $\bsd=\bsF$ are plotted in \fref{fig: 2dflow} (d). The black and green lines are based on \eref{max TUR} and correspond to $\widehat{\sigma^\tot}=(j_d^\hk+j_d^\ex)^2/\calD_d$ and $(j_d^\hk-j_d^\ex)^2/\calD_d$, respectively. 
The black line represents the estimation based on the conventional TUR, \eref{TUR}, and the lower bound $\widehat{\sigma^\tot}$ is getting tighter in the region $T'\gg T$. This is because the coefficient is asymptotically proportional to the thermodynamic force in the limit $T'\rightarrow \infty$. By contrast, this bound gets looser for $T'<T$, and the precision $\widehat{\sigma^\tot}/\sigma^\tot$ takes the lowest value $0$ around $T'= T/4$. This is because the power vanishes at this point (see \fref{fig: 2dflow} (b)) and cannot reproduce the nonzero EP. This situation has been confirmed for the stopping force in \ccite{Kamijima-HOEB}. Moreover, since the product of the currents $j_d^\hk j_d^\ex$ is negative in this temperature domain, the green line surpasses the black line and the estimation is better than the conventional TUR. Note that this difference {corresponds to whether the current state has a higher potential} energy than the steady state or not.

We apply the inequalities \eqref{respective TUR} to the housekeeping and excess EP rates. Figure \ref{fig: 2dflow} (d) shows that we can estimate roughly $80\%$ of the housekeeping EP rate and $20\%$ of the excess EP rate (the red line) by using directly measurable quantities. The precision, $\widehat{\sigma^\hk}/\sigma^\hk$ and $\widehat{\sigma^\ex}/\sigma^\ex$, are independent of $T'$, because the lower bounds have the same $T'$-dependency as the true value. 

\begin{figure}[t]
    \centering
    \includegraphics[width=0.8\linewidth]{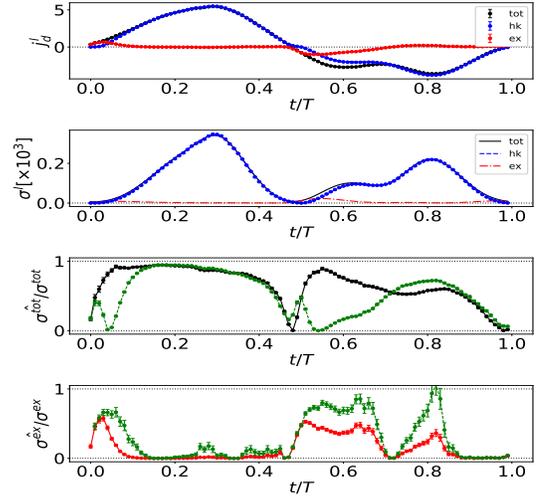}
    \caption{\label{fig: rocking} (a) The calculation of the currents with $\bsd(x)=1$. The points represent the calculation based on the Langevin equation, taking the sample averages. The lines represent the calculation based on solving the initial-value problem of the Fokker-Planck equation. They agree with each other quite well. (b) The calculation of the EP rates. Since we know the instantaneous steady state but not the periodic steady state, we can estimate only the housekeeping EP rate based on the Langevin equation. The excess EP rate is one order of magnitude smaller than the housekeeping EP rate. (c) The estimation of the total EP rate based on \eref{max TUR} using the quantities obtained only by the Langevin equation. The black (green) line represents the estimation with the positive (or negative) sign,  $j_d^\hk+j_d^\ex$ (or $j_d^\hk-j_d^\ex$). (d) The estimation of the excess EP rate based on \eref{projective TUR} (green line) and \eref{respective TUR} (red line). The parameters used in calculation are $D = 0.1, T=0.7, a=0.8, R_0 = 6.0$. In the Langevin equation, we take the averages over $10^4$ trajectories. In the Fokker-Planck equation, we use the resolution of $2^{10}$ elements for the Fourier decomposition.
    }
\end{figure}

\textit{Application to a rocking ratchet.---}
We next consider a Brownian particle in a ratchet potential.
The drift term and the noise strength are given by $F=-\del_x U(x)+R(t)$ and $G=\sqrt{D}$ ($\mu=1$), respectively. The rocking force $R(t)$ is periodic in time as $R(t+\calT)=R(t)$ with a period $\calT(>0)$. The potential $U(x)$ is periodic in space as $U(x+L)=U(x)$. 
{Note that, even if the rocking force has no bias on average, a finite particle current is obtained if the potential breaks the left-right symmetry.}
The dynamics in the infinite domain reduces to the one in the reduced domain $[0, L]$, i.e., $p(x+L,t)=p(x,t)$ and $\int_0^L p(x, t)=1$. 
This model has also been examined in \ccite{Kim-Hollerbach2022}. 
A crucial feature of this model is that we can easily calculate the instantaneous steady state $p^\ssrm(x, t)$, {in contrast to the long-time time-periodic state}. The distribution is a modified canonical distribution which takes into account the spatial periodicity \cite{Reimann2002} (see Sec. IV of Supplemental Material).

As in \ccite{Kim-Hollerbach2022}, we consider a saw-tooth potential such that $U(x)=U_0 x/\alpha L$ for $0<x<\alpha L$ ($0<\alpha<1$) and $U(x)=U_0 (x-L)/(1-\alpha) L$ for $\alpha L<x<L$. The rocking force is assumed to be sinusoidal, $R(t)=R_0\sin(2\pi t/\calT)$.

We numerically calculate the currents and the EPs by sampling trajectories of the Langevin dynamics. We wait for a sufficiently long time that the system relaxes to its time-periodic state. Since we know the instantaneous steady state, we can calculate $j_d=\bkt{d\circ \dot{x}},\ j_d^\hk=\bkt{d\ v^\ssrm},\  j_d^\ex=j_d-j_d^\hk$ and $\sigma^\hk=\bkt{v^\ssrm\circ \dot{x}}$ by taking the sample average, analogous to an experimental measurement. The Stratonovich product $\circ\dot{x}$ is converted to It\^{o} product to achieve a faster convergence in the calculation.

On the other hand, $\sigma^\tot$ and $\sigma^\ex$ are inaccessible in this method because {we do not have an explicit expression for} the time-periodic state. Therefore, we evaluate them by applying our TUR, \eref{max TUR} and \eref{projective TUR}, to this model. For this evaluation, we exactly calculate $\sigma^\tot$ and $\sigma^\ex$ by solving the Fokker-Planck equation, which is numerically much harder especially at low temperatures \cite{Kim-Hollerbach2022}.

The currents and EP rates are plotted in \fref{fig: rocking} (a) and (b). The currents change their signs around $t=\calT/2$ in accordance with the rocking force $R(t)$. The excess EP rate is small compared to the others, which implies that the time-periodic state is similar to the instantaneous steady state and the unidirectional transport is driven mainly by the nonconservative force itself.

In addition, in the instantaneous steady state, the mean local velocity takes a step-like form throughout the period due to the rocking force and saw-tooth potential (see Sec. IV of Supplemental Material). This implies that the coefficient roughly satisfies $d\propto e^\hk$ (\eref{hk coeff}) for the particle current with $d(x)=1$, leading to a small excess current. Nevertheless, the excess components (the current and the EP rate) become large at $t/\calT=0, 0.5$, because the housekeeping components have to vanish by definition, regardless of the nonzero total components stemming from the time-dependent driving.

Figure \ref{fig: rocking} (c) illustrates the estimation of the total EP based on \eref{max TUR}. The black and green lines correspond to $\widehat{\sigma^\tot}=(j_d^\hk+j_d^\ex)^2/\calD_d$ and  $\widehat{\sigma^\tot}=(j_d^\hk-j_d^\ex)^2/\calD_d$, respectively. Since the housekeeping and excess currents have the same sign in the almost whole region (see \fref{fig: rocking} (a)), there is no significant advantage to use \eref{max TUR} over the conventional TUR \eqref{TUR}. Note that this depends on the behavior of the particle current and the estimation may be improved for another choice of the current. 

As for the excess EP, we apply inequalities \eqref{projective TUR} and \eqref{respective TUR} to the estimation (see \fref{fig: rocking} (d)).
The black and green lines correspond to $\widehat{\sigma^\ex}=(j_d^\ex)^2/(\calD_d-(j_d^\hk)^2/\sigma^\hk), (j_d^\ex)^2/\calD_d$, respectively. By definition, the green line surpasses the red line. In the latter half of the period, the estimation is relatively good and culminates around $t=0.8T$, achieving almost 100$\%$ precision. By contrast, the estimation gets worse in the first half of the period, because the excess current almost vanishes in this regime (see \fref{fig: rocking} (a)). This can be resolved by using a low-frequency current coefficient (see Sec. IV of Supplemental Material).

\textit{Discussion.---}
We have extended the thermodynamic uncertainty relation to the framework of steady-state thermodynamics, which is the projective TUR \eqref{projective TUR}. We show that the excess-housekeeping decomposition can also be applied to currents as well as the entropy production, and that the respective components satisfy a TUR both separately and together. The newly introduced currents, $j_d^\hk,j_d^\ex$, contain information on the housekeeping and excess entropy productions. We clarify the condition when it is possible to measure these currents in the same way as the total current, Eqs. \eqref{hk coeff} and \eqref{ex coeff}. Our TUR yields various corollaries that can be used according to the experimental restriction.

We illustrated our TUR in two paradigmatic examples. In the first example, the two-dimensional flow offers a clear physical interpretation of the housekeeping and excess currents. This model is expected to be realizable by using optical tweezers \cite{Wu-Rongxin2009}. 
The second example, the rocking ratchet, depicts the situation when we can estimate the excess entropy production, taking advantage of the knowledge of the steady state. The rocking ratchet can be realized by nanofluidic circuitry \cite{Michael-Christian2018}.
In this example, we can divide the particle current into the excess and housekeeping parts, and the relative size of these contributions can tell us whether the physical mechanism mainly driving the transport is attributed to the nonconservative force or the time-dependence of the system state. Likewise, detailed information about the currents can help us to understand the dynamics of nonequilibrium processes and to optimize thermodynamic machines such as ratchets and heat engines.


\textit{Acknowledgement.}
T.K. is supported by World-leading Innovative Graduate Study Program for Materials Research, Industry, and Technology (MERIT-WINGS) of the University of Tokyo.
{S.I. is supported by JSPS KAKENHI Grants No.19H05796, No.21H01560, and No.22H01141, and UTEC-UTokyo FSI Research Grant Program.
A.D. is supported by JSPS KAKENHI Grant Numbers 19H05795 and 22K13974.}
T.S. is supported by JSPS KAKENHI Grant Numbers JP16H02211 and JP19H05796. T.S. is also supported by Institute of AI and Beyond of the University of Tokyo.

\clearpage
\begin{widetext}
\begin{center}
{\large \bf Supplemental Material for  \protect \\ 
  ``Thermodynamic Uncertainty Relations for Steady-State Thermodynamics'' }\\
\vspace*{0.3cm}
Takuya Kamijima$^{1}$, Sosuke Ito$^{2}$, Andreas Dechant$^{3}$  and Takahiro Sagawa$^{1,4}$ \\
\vspace*{0.1cm}
{$^{1}$Department of Applied Physics, The University of Tokyo, 7-3-1 Hongo, Bunkyo-ku, Tokyo 113-8656, Japan}\\
{$^{2}$Department of Physics, The University of Tokyo, 7-3-1 Hongo, Bunkyo-ku, Tokyo 113-0033, Japan}\\
{$^{3}$Department of Physics No. 1, Graduate School of Science, Kyoto University, Kyoto 606-8502, Japan}\\
{$^{4}$Quantum-Phase Electronics Center (QPEC), The University of Tokyo, 7-3-1 Hongo, Bunkyo-ku, Tokyo 113-8656, Japan}
\end{center}
\end{widetext}

\section{Generalization and extension}
We assume that thermodynamic forces $D^{-1}\bsv^\alpha(\alpha=1, \cdots, n)$ satisfy the following two conditions, (i) the constraint condition: $\sum_{\alpha=1}^n\bsv^\alpha=\bsv$ and (ii) the orthogonal condition $\bkt{D^{-1}\bsv^\alpha,D^{-1}\bsv^\beta}\propto\delta_{\alpha,\beta}$. The corresponding current and EP are defined as $j_d^\alpha=\bkt{\bsd,D^{-1}\bsv^\alpha}$ and $\sigma^\alpha:=||D^{-1}\bsv^\alpha||^2$ respectively.

Including the normalized vector fields $\bse^\alpha=D^{-1}\bsv^\alpha/||D^{-1}\bsv^\alpha||$ in the basis, \eref{projective TUR} can be generalized to
\begin{align}
    \sum_{\alpha=1}^n \frac{(j_d^\alpha)^2}{\sigma^\alpha}\leq \calD_d
\end{align}
in the same way as \eref{derivation of projective TUR}. In \eref{projective TUR}, the chosen basis includes the housekeeping and excess vector fields $(\alpha=\hk,\ex)$, and the thermodynamic forces already satisfy the constraint condition. Therefore, even though the lower bound can be tightened by adding other components of the basis, $\bse^\alpha$, the components have no information about the norm of the corresponding thermodynamic force, making $j_d^\alpha$ and $\sigma^\alpha,$ undefined.

For the thermodynamic forces and arbitrary signs $\chi^\alpha=\pm1\ (\alpha=1, \cdots, n)$, the following inequality holds: 
\begin{align}
    &\left\langle\bsd,D^{-1}\sum_{\alpha=1}^n\chi^\alpha\bsv^\alpha\right\rangle\nonumber\\
    &\leq\bkt{\bsd,\bsd}\left\langle D^{-1}\sum_{\alpha=1}^n\chi^\alpha\bsv^\alpha,D^{-1}\sum_{\alpha=1}^n\chi^\alpha\bsv^\alpha\right\rangle\nonumber\\
    &=\bkt{\bsd,\bsd}\bkt{D^{-1}\bsv, D^{-1}\bsv},
\end{align}
where the Cauchy-Schwartz inequality is used in the first line. From this inequality, \eref{max TUR} can be generalized  to
\begin{align}
    \sigma^\tot\geq\frac{1}{\calD_d}
    \max_{\chi^\alpha}\left(\sum_{\alpha=1}^n \chi^\alpha j_d^\alpha\right)^2.
\end{align}
The equality is achieved if and only if the coefficient satisfies $\bsd\propto D^{-1}\sum_{\alpha=1}^n \chi^\alpha \bsv^\alpha$. For the fixed coefficient, the best estimation of the dissipation is obtained for $\chi^\alpha$ which makes the signs of $\chi^\alpha j_d^\alpha$ aligned.
{The equality condition of \eref{max TUR} is obtained by setting $\alpha=\hk,\ex$. For the case of $j_d^\hk j_d^\ex>0$, the equality condition is $\bsd={j_d^\hk}D^{-1}\bsv/{\sigma^\hk}\propto D^{-1}\bsv$ and the current coefficient is proportional to the thermodynamic force \cite{Gingrich-Horowitz2016}. By contrast, for the case of $j_d^\hk j_d^\ex<0$, the above relation is replaced by  $\bsd=-{j_d^\hk}D^{-1}(\bsv-2\bsv^\ssrm)/{\sigma^\hk}\propto D^{-1}\bsv^\dagger$ (with $\bsv^\dagger:=\bsv-2\bsv^\ssrm$), which means that the current coefficient is proportional to the thermodynamic force in the dual dynamics \cite{Dechant-Sasa-continuous}.}

{While the single coefficient $d$ is used in inequalities \eqref{projective TUR} and \eqref{respective TUR}, two coefficients $d$ and $d'$ can be utilized to assess $\sigma^\hk$ and $\sigma^\ex$ respectively.}
\eref{respective TUR} can be modified for two coefficients as
\begin{align}
    \sigma^\tot\geq
    \frac{\left(j_{d}^\hk\right)^2}{D_{d}}
    +\frac{\left(j_{d'}^\ex\right)^2}{D_{d'}}.
\end{align}
This TUR can be applied for the case when the kind of coefficient is restricted by the experimental condition. The equality is obtained if and only if $\bsd\propto\bse^\hk,\bsd'\propto\bse^\ex$ is satisfied. {Therefore, if we set $\bsd=\bsf\propto D^{-1}\bsv^\ssrm$ and $\bsd'=-\bsna U\propto D^{-1}(\bsv-\bsv^\ssrm)$ in the example of the two-dimensional Brownian motion, we can completely estimate the housekeeping and excess EP rate. Therefore, we can estimate all the EP rates by taking measurements for these particular two currents.}
Note that \eref{projective TUR} is still tighter as long as we focus on the estimation with the single coefficient.
Moreover, adding small coefficient $\delta d$ such that $\delta d\propto\bse^\ex,\delta d'\propto\bse^\hk$, \eref{projective TUR} can be extended to
\begin{align}
    \sigma^\tot\geq
    \frac{\left(j_{d}^\hk\right)^2}{D_{d+\delta d}}
    +\frac{\left(j_{d'}^\ex\right)^2}{D_{d'+\delta d'}}.
\end{align}

\section{Markov jump processes}
For the Langevin dynamics, the orthogonal condition \eqref{orthogonal condition} is satisfied because the surface term vanishes due to its boundary condition. By contrast, this condition does not hold for Markov jump processes, and neither \eref{projective TUR} nor \eqref{max TUR} can be derived. On the other hand, as the counterpart of \eref{respective TUR}, the following TURs hold:
\begin{align}
    \label{Markov jump: respective TUR}
    \sigma^\hk\geq c_0\frac{\left(j_d^\hk\right)^2}{\tilde{D}_d},\ 
    \sigma^\ex\geq c_0\frac{\left(j_d^\ex\right)^2}{\tilde{D}_d}.
\end{align}
Contrary to \eref{respective TUR}, the constant $c_0=0.896\dots$ is required and the variance of the current $\tilde{D}_d$ is also different. It is nontrivial whether the average and variance of the current are directly measurable or not. Nonetheless, we see that these take a simple form (\eref{QD: hk current}\eqref{QD: ex current}\eqref{QD: HS variance}) for the example of the quantum dot mentioned below.

Compared to the conventional TUR, \eref{TUR},
\begin{align}
    \label{Markov jump: sign condition}
    \frac{(j_d)^2}{\calD_d}
    =\frac{1}{c_0}\frac{\tilde{D}_d}{\calD_d}
    \left(c_0\frac{\left(j_d^\hk\right)^2}{\tilde{D}_d}+c_0\frac{\left(j_d^\ex\right)^2}{\tilde{D}_d}\right)
    +\frac{2j_d^\hk j_d^\ex}{\calD_d}
\end{align}
means that, roughly speaking, it is better to estimate the total dissipation by decomposing it into the housekeeping and excess parts when the product of the currents $j_d^\hk j_d^\ex$ is negative.

\subsection{Derivation}
We assume that the system is coupled to multiple reservoirs and its discrete state $x$ obeys the master equation:
\begin{align}
    \dot{p}_{x}
    =\sum_\nu\sum_{x'}R_{xx'}^\nu p_{x'}
    =\sum_\nu\sum_{x'}j_{xx'}^\nu,
\end{align}
where $j_{xx'}^\nu:=R_{xx'}^\nu p_{x'}-R_{x'x}^\nu p_{x}=K_{xx'}^\nu-K_{x'x}^\nu$ is the probability flow from the state $x'$ to $x$ driven by the reservoir $\nu$ $(K_{xx'}^\nu:=R_{xx'}^\nu p_{x'})$.
In the steady state, this satisfies the balanced condition $\sum_\nu\sum_{x'}{j^\ssrm}_{xx'}^\nu=0$.
The EP rates are given by \cite{Three-faces-Master-equation,3DFT-Esposito2010}
\begin{align}
    \label{Markov jump: total EP}
    \sigma^\tot&=\sum_\nu\sum_{x\neq x'}
    K_{xx'}^\nu\ln\frac{K_{xx'}^\nu}{K_{x'x}^\nu},\\
    \label{Markov jump: hk EP}
    \sigma^\hk&=\sum_\nu\sum_{x\neq x'}
    K_{xx'}^\nu\ln\frac{K_{xx'}^\nu}{\tilde{K}_{xx'}^\nu},
\end{align}
and
\begin{align}
    \label{Markov jump: ex EP}
    \sigma^\ex=\sum_\nu\sum_{x\neq x'}
    K_{xx'}^\nu\ln\frac{K_{xx'}^\nu}{\tilde{K}_{x'x}^\nu}.
\end{align}
We define $\tilde{K}$ as $\tilde{K}_{xx'}^\nu:=\tilde{R}_{xx'}^\nu p_{x'}$ with the dual transition rate $\tilde{R}_{xx'}^\nu:=R_{x'x}^\nu p^\ssrm_{x}/p^\ssrm_{x'}$.

Alhough the mean local velocity is not defined conventionally for Markov jump processes, we introduce it with the definition $v_{xx'}^\nu:=j_{xx'}^\nu/p_{x'}$ in analogy with the Langevin dynamics. Dividing the probability flow $j_{xx'}$ by the probability $p_{x'}$, the mean local velocity captures the net current of the transition. On the contrary to the probability flow, the mean local velocity does not satisfy the skewed symmetry, i.e., $v_{xx'}^\nu\neq v_{x'x}^\nu$.

The total current average is expressed as
\begin{align}
    \label{Markov jump: def of current}
    j_d(t)
    &=\sum_\nu\sum_{x\neq x'}d_{xx'}^\nu K_{xx'}^\nu
    =\frac{1}{2}\sum_\nu\sum_{x\neq x'}d_{xx'}^\nu j_{xx'}^\nu\nonumber\\
    &=\frac{1}{2}\sum_\nu\sum_{x\neq x'}d_{xx'}^\nu v_{xx'}^\nu p_{x'}.
\end{align}
We define the housekeeping and excess current averages as
\begin{align}
    \label{Markov jump: def of hk current}
    j_d^\hk(t)
    :&=\frac{1}{2}\sum_\nu\sum_{x\neq x'}d_{xx'}^\nu 
    (K_{xx'}^\nu-\tilde{K}_{xx'})\nonumber\\
    &=\frac{1}{2}\sum_\nu\sum_{x\neq x'}d_{xx'}^\nu {v^\ssrm}_{xx'}^{\nu} p_{x'}
\end{align}
and 
\begin{align}
    \label{Markov jump: def of ex current}
    j_d^\ex(t)
    :&=\frac{1}{2}\sum_\nu\sum_{x\neq x'}d_{xx'}^\nu 
    (K_{xx'}^\nu+\tilde{K}_{xx'})\nonumber\\
    &=\frac{1}{2}\sum_\nu\sum_{x\neq x'}d_{xx'}^\nu 
    ({v^\ssrm}_{xx'}^{\nu}-v_{xx'}^\nu)p_{x'}
\end{align}
respectively. Then, the current decomposition $j_d=j_d^{\textrm{hx}}+j_d^\ex$ holds by definition. These current averages have almost the same form as their counterparts in Langevin dynamics and a simple calculation confirms that these coincide in the continuous limit.
In addition, we define both the housekeeping and excess current variances as
\begin{subequations}
\begin{align}
    \tilde{D}_d:&=\frac{1}{4}\sum_\nu\sum_{x\neq x'}(d_{xx'}^\nu)^2
    (K_{xx'}^\nu+\tilde{K}_{xx'}^\nu)\\
    &=\frac{1}{4}\sum_\nu\sum_{x\neq x'}(d_{xx'}^\nu)^2
    (K_{xx'}^\nu+\tilde{K}_{x'x}^\nu).
\end{align}
\end{subequations}

We are now ready to derive \eref{Markov jump: respective TUR}.
\begin{align}
    \sigma^\hk\tilde{D}_d
    =\sum_\nu&\sum_{x\neq x'}\left(K_{xx'}^\nu
    \ln\frac{K_{xx'}^\nu}{\tilde{K}_{xx'}^\nu}
    +\tilde{K}_{xx'}^\nu-K_{xx'}^\nu\right)\nonumber\\
    &\cdot\frac{1}{4}\sum_\nu\sum_{x\neq x'}(d_{xx'}^\nu)^2
    (K_{xx'}^\nu+\tilde{K}_{xx'}^\nu)\nonumber\\
    \geq c_0 &\sum_\nu\sum_{x\neq x'}
    \frac{(K_{xx'}^\nu-\tilde{K}_{xx'}^\nu)^2}
    {K_{xx'}^\nu+\tilde{K}_{xx'}^\nu}\nonumber\\
    &\cdot\frac{1}{4}\sum_\nu\sum_{x\neq x'}(d_{xx'}^\nu)^2
    (K_{xx'}^\nu+\tilde{K}_{xx'}^\nu)
    \nonumber\\
    \geq c_0 &\left(\frac{1}{2}\sum_\nu\sum_{x\neq x'}
    d_{xx'}^\nu(K_{xx'}^\nu-\tilde{K}_{xx'}^\nu)\right)^2\nonumber\\
    =c_0&\left(j_d^\hk\right)^2
\end{align}
The derivation for the excess part follows in exactly the same way. We used an equality $\sum_x K_{xx'}^\nu=\sum_x \tilde{K}_{xx'}^\nu=0$, an inequality 
\begin{align}
    \label{basic ineq c=0.896}
    a\ln\frac{a}{b}+b-a\geq \frac{c_0(a-b)^2}{a+b}
\end{align}
($a,b>0, c_0=0.896\dots$) \cite{Shiraishi-Funo-Saito,Shiraishi-Saito-Tasaki2016}, and the Cauchy-Schwaltz inequality in each transformation.

\subsection{The necessity of $c_0$}
\begin{figure}
    \centering
    \includegraphics[width=1.0\linewidth]{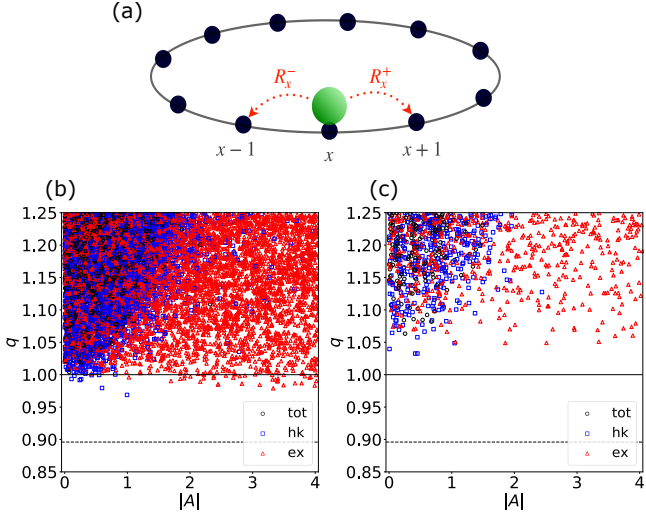}
    \caption{\label{fig: MJ_RW}
    (a) The sketch of the random walk. A single particle is hopping among sites driven by a heat reservoir.
    (b)(c)The test of the short-time TUR for each EP rate in the $1$-dimensional random walk (\eref{1dRW: tot TUR}\eqref{1dRW: HS TUR}). The parameters of the model are chosen at random from $A\in[-4, 4],\ R_x\in[0, 1],\ d_x\in[-1, 1]$. The probability distribution of the system $p_x$ is also at random. The sample size is $10^5$. The system size is $M=3$ for (b), and $M=5$ for (c).
    }
\end{figure}
We see the constant $c_0=0.896\dots$ is required for Markov jump processes in the simple one-dimensional random walk model. There are $M$ sites in the one direction with the periodic boundary condition, and a single particle is hopping among them, being driven by a heat reservoir (see \fref{fig: MJ_RW} (a)). We denote the transition rate to the forward (backward) direction at site $x(=1,2,\dots,M)$ as $R_{x,+}\ (R_{x,-})$. We assume the local detailed balance condition for this system, $\frac{R_{x,+}}{R_{x+1,-}}=e^A$, with the thermodynamic force $A$. Thus, the transition rate is written by $R_{x,+}:=R_xe^{A/2},\ R_{x,-}:=R_{x-1}e^{-A/2}$.
The current coefficient obeys the skewed symmetry:
\begin{align}
    d_{xx'}=\begin{cases}
            d_{x-1} & (x'=x-1)\\
            -d_x & (x'=x+1)\\
            0 & (\textrm{otherwise})
            \end{cases}.
\end{align}

In this setup, we examine the TURs
\begin{align}
    \label{1dRW: tot TUR}
    q:&=\frac{\sigma^\tot {D}_d}
    {\left(j_d\right)^2}\geq1\\
    \label{1dRW: HS TUR}
    q:&=\frac{\sigma^l \tilde{D}_d}
    {\left(j_d^l\right)^2}\geq c_0
    \ (l=hk,\ex)
\end{align}
by plotting $q$ for many systems with different probability distributions, transition rates, and current coefficients.
In \fref{fig: MJ_RW} (b), the lower bound $q\geq1$ is violated for some blue and red points, meaning that \eref{Markov jump: respective TUR} does not hold without the constant $c_0$. At the same time, there are few points near the bound $q\geq c_0$, which suggests that the quality condition is hard to satisfy.

Since the conventional TUR \eqref{TUR} is satisfied in the equilibrium limit, the lower bound $q\geq1$ gets tighter for small thermodynamics forces. The blue plot behaves similarly to the black plot because the housekeeping EP captures the violation of the detailed balance condition. On the other hand, the excess EP comes from the deviation from the steady state. Therefore, the red points are widely distributed and the tightness of the lower bound is almost independent of $A$.

The samples close to the lower bound get scarce when we increase the size of the system (see \fref{fig: MJ_RW} (c)). This suggests that achieving equality requires the state and coefficients to be tuned commensurately and this tuning gets extremely hard for large systems. Therefore, the lower bound $q\geq$ is practically never broken for sufficiently large systems, and the same TUR as the Langevin dynamics, \eref{respective TUR}, can be applied.

\subsection{Example: quantum dot}
The EP is estimated using TUR \eqref{Markov jump: respective TUR} in the model of the quantum dot. The number of levels is one, and we denote state 1 when the particle is occupied and state 0 when it is empty. The transition rate of the heat reservoir $\nu(=\textrm{h,c})$ is given by the Fermi golden rule as $R_{10}^\nu=\Gamma_0 f_\nu,\ R_{01}^\nu=\Gamma_0(1-f_\nu)\ (f_\nu:=1/(1+e^{\beta_\nu(E-\mu_\nu)}))$ \cite{Benenti-Casati-Saito-Whitney2017, Esposito-Lindenberg-Broeck2009-QD}. $E$ is the energy of the level and $\mu_\nu$ is the chemical potential of the heat reservoir $\nu$. The coupling strength to the heat reservoir is assumed to be the same $\Gamma_0$, which is sufficiently small that the sequential tunneling approximation holds.

The steady state is easily calculated:
\begin{align}
    p_0^\ssrm=\frac{R_{01}}{R_{01}+R_{10}},\ 
    p_1^\ssrm=\frac{R_{10}}{R_{01}+R_{10}}
    =\frac{1}{2}(f_{\textrm{h}}+f_{\textrm{c}}).
\end{align}
The distinguished feature of 2-state systems is that the overall dual transition rate matches the original transition rate, that is, 
\begin{align}
    \tilde{R}_{xx'}=\sum_\nu\tilde{R}_{xx'}^\nu=\sum_\nu\frac{{R}_{x'x}^\nu p_{x}^\ssrm}{p_{x'}^\ssrm}={R}_{xx'}.
\end{align}

\begin{figure}
    \centering
    \includegraphics[width=1.0\linewidth]{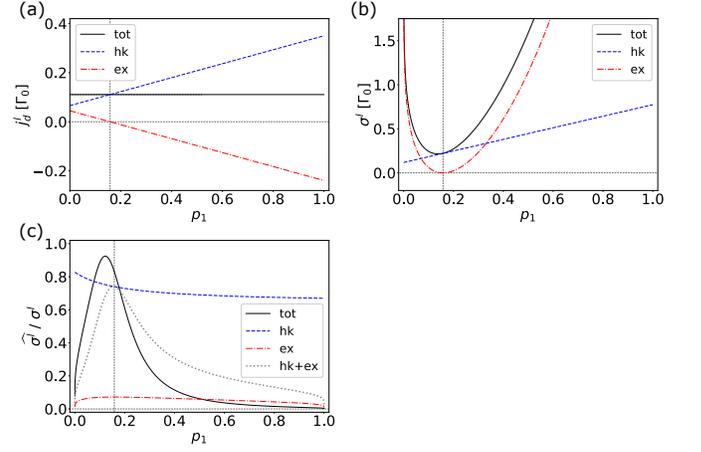}
    \caption{\label{fig: MJ_QD}
    (a) The averages of the particle current in the quantum dot model. All of the averages are linear in terms of $p_1$. The vertical dot line depicts the steady state, $p_1=p_1^\ssrm$. (b) Each EP rate. In the steady state, the excess EP rate vanishes and the housekeeping EP rate coincides with the total one. (c) The estimation of each EP rate based on \eref{TUR}\eqref{Markov jump: respective TUR} using the particle current. $\widehat{\sigma^l}$ denotes the estimation of $\sigma^l\ (l=\tot,\hk,\ex)$. The gray line represents the estimation with $\widehat{\sigma^\tot}=c_0[(j_d^\hk)^2+(j_d^\ex)^2]/\tilde{D}_d$. The parameters used in calculation are $\beta_\textrm{h}(E-\mu_\textrm{h})=1.0,\beta_\textrm{c}(E-\mu_\textrm{c})=3.0$.
    }
\end{figure}
As a current, consider a particle current through the system in the direction from the high-temperature heat reservoir to the low-temperature heat reservoir. The coefficient is given by $d_{xx'}^\nu=M_{xx'}^\nu$ with
\begin{align}
    M^{\textrm{h}}_{xx'}=-M^{\textrm{c}}_{xx'}=\frac{(N_x-N_{x'})}{2}.
\end{align}
$N_x$ is the number of particles of the system in state $x$, where $N_0=0,N_1=1$. Since we are dealing with non-stationary conditions here, we need to consider the exchange of particles with both heat reservoirs. 
Computing each current average based on \eref{Markov jump: def of current}-\eqref{Markov jump: def of ex current},
\begin{align}
    \label{QD: tot current}
    j_d&=\sum_\nu M_{10}^\nu(R_{10}^\nu p_0-R_{01}^\nu p_1)=\frac{\Gamma_0}{2}(f_h-f_c)\\
    \label{QD: hk current}
    j_d^\hk
    &=\frac{1}{2}\sum_\nu M_{10}^\nu
    \left[(R_{10}^\nu p_0-\tilde{R}_{10}^\nu p_0)
    -(R_{01}^\nu p_1-\tilde{R}_{01}^\nu p_1)\right]\nonumber\\
    &=\frac{\Gamma_0}{4}\left(\frac{p_0}{p_0^\ssrm}
    +\frac{p_1}{p_1^\ssrm}\right)(f_h-f_c)\\
    \label{QD: ex current}
    j_d^\ex
    &=j_d^\tot-j_d^\hk\nonumber\\
    &=\frac{\Gamma_0}{4}\left(\left(1-\frac{p_0}{p_0^\ssrm}\right)
    +\left(1-\frac{p_1}{p_1^\ssrm}\right)\right)(f_h-f_c)
\end{align}
These are plotted in \fref{fig: MJ_QD} (a). It can be seen that the current mean of the total system is always constant, independent of the state probability $p_x$. This is due to the two-state nature and the property $M^{\textrm{h}}_{xx'}=-M^{\textrm{c}}_{xx'}$. The housekeeping and excess current averages are the current averages of the total system multiplied by $({p_0}/{p_0^\ssrm}+{p_1}/{p_1^\ssrm})/2$ and $((1-{p_0}/{p_0^\ssrm})+(1-{p_1}/{p_1^\ssrm)})/2$, respectively. The housekeeping current average is always positive and the excess current average changes its sign at $p_1=p_1^\ssrm$.

On the other hand, each current variance is computed as 
\begin{align}
    \label{QD: variance}
    \calD_d
    &=\frac{1}{8}\sum_\nu(R_{10}^\nu p_0+R_{01}^\nu p_1)\\
    \label{QD: HS variance}
    \tilde{D}_d&=
    \frac{1}{16}\sum_\nu
    \left[(R_{10}^\nu p_0+\tilde{R}_{10}^\nu p_0)
    +(R_{01}^\nu p_1+\tilde{R}_{01}^\nu p_1)\right]\nonumber\\
    &=\frac{1}{8}\sum_\nu(R_{10}^\nu p_0+R_{01}^\nu p_1)
    =\calD_d
\end{align}
and is the same value. This is also due to the special properties of the two states and $M^{\textrm{h}}_{xx'}=-M^{\textrm{c}}_{xx'}$, where the variance generally takes different values.

In this model, the EP rates calculated by \eref{Markov jump: total EP}-\eqref{Markov jump: ex EP} are shown in \fref{fig: MJ_QD} (b). The excess EP rate takes a minimum value of $0$ at the steady state $p_1=p_1^\ssrm$ and increases rapidly away from it. The housekeeping EP rate is linear and monotonically increasing because $\ln(K_{x'x}^\nu/\mathcal{K}_{x'x}^\nu)$ is independent of $p_1$. Also, from a simple calculation, it can be shown that this slope increases as the thermal force $A=\beta_\textrm{c}(E-\mu_\textrm{c})-\beta_\textrm{h}(E-\mu_\textrm{h})$ is larger. Reflecting these behaviors, the EP rate for the total system reaches a finite minimum at $p_1$ which is slightly smaller than $p_1^\ssrm$.

Finally, the EP rate estimations given by the TUR \eqref{Markov jump: respective TUR} are plotted in \fref{fig: MJ_QD} (c). The accuracy of the EP rate for the total system tends to have a maximum value of about $90\%$ near the minimum of the EP rate, and the accuracy decreases as one moves away from the minimum. For the excess EP rate, the overall accuracy is less than $10\%$, but the accuracy improves as the steady state is approached. On the other hand, for the housekeeping EP rate, the overall accuracy is high, and as $p_1$ increases, the accuracy gradually decreases. The gray line in the figure shows the estimated EP rate for the entire system from the lower bounds of the housekeeping and excess EP rates. From \eref{Markov jump: sign condition}, it is better to decompose and estimate the housekeeping and excess components approximately when the product of the currents $j_d^\hk j_d^\ex$ becomes negative. In fact, for $p_1>p_1^\ssrm$ where the product of the currents is negative (see \fref{fig: MJ_QD}(a)), the gray line is located above the black line except near its boundary, and the EP rate for the total system is estimated more accurately.

\section{Details of the example: 2-dimensional flow}
In this section, we present some calculations and figures to supplement the first example in the main text.
Since $D$ is proportional to the identity matrix, the coefficient proportional to the nonconservative force, $\bsd=\bsd_\theta:=d_\theta[-x_2, x_1]^T$, satisfies \eref{hk coeff}, and the coefficient proportional to the conservative force, $\bsd=\bsd_r:=-d_r[x_1, x_2]^T$, satisfies \eref{ex coeff}. Here, we consider coefficients including only $\bsd_\theta,\bsd_r$, but note that this does not mean that all coefficients can be expanded by them.

\begin{figure}
    \centering
    \includegraphics[width=0.8\linewidth]{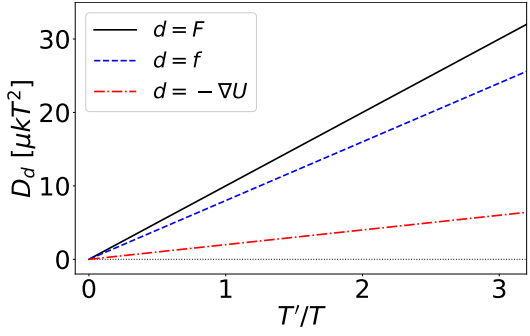}
    \caption{\label{fig: 2dflow_variance}
    The variance of the power exerted on the system in the 2-dimensional flow model. Due to the decomposition of the coefficient, \eref{2Dflow: decomp of fluctuation}, the black line agrees with the sum of the blue and red lines. The parameters used in calculation are $\mu=1.0, k=1.0, \kappa=2.0, T=1.5$.
    }
\end{figure}

The current averages with $\bsd=\bsd_\theta+\bsd_r$ are given by
\begin{align}
    \label{2Dflow: hk coeff}
    j_d^\hk&=j_{d_\theta}^\hk=j_{d_\theta}
    =\frac{2\kappa T'd_\theta}{k}
\end{align}
and
\begin{align}
    \label{2Dflow: ex coeff}
    j_d^\ex&=j_{d_r}^\ex=j_{d_r}
    =2 T'\left(1-\frac{T}{T'}\right)d_r.
\end{align}
While $J_{D}^\hk$ has the same sign regardless of $T'$, $J_{D}^\ex$ switches positive and negative at $T'=T$. The current variances with $\bsd=\bsd_\theta,\ \bsd_r$ are computed as
\begin{align}
    D_{d_\theta}=\frac{2 TT'{d_\theta}^2}{k},\ \textrm{and }
    D_{d_r}=\frac{2 TT'{d_r}^2}{k}
\end{align}
respectively. Since $\bsd_\theta^T\bsd_r=0$ (and $D$ is proportional to the identity matrix), 
the current with $\bsd=\bsd_\theta+\bsd_r$ satisfies the decomposition of the variance as well:
\begin{align}
    \label{2Dflow: decomp of fluctuation}
    \calD_d=D_{d_\theta}+D_{d_r}.
\end{align}
For the coefficient $\bsd=\bsF$, the averages are plotted in \fref{fig: 2dflow} (b) and the variance is plotted in \fref{fig: 2dflow_variance}.

Next, we calculate the EP rates. For the current coefficient,
\begin{align}
    \bsd=D^{-1}\bsv
    &=\frac{1}{T}(-\bsna U+\bsf-T\bsna\ln p)\nonumber\\
    &=\frac{\kappa}{T}\bsd_\theta+\frac{k}{T}
    \left(1-\frac{T}{T'}\right)\bsd_r,
\end{align}
the housekeeping (excess) EP rate is given as its housekeeping (excess) current due to the orthogonal condition \eqref{orthogonal condition}:
\begin{align}
    \sigma^\hk&=\frac{2\kappa^2}{k}\frac{T'}{T},
\end{align}
and
\begin{align}
    \sigma^\ex&=\frac{2\kappa^2}{k}\frac{T'}{T}\left(\frac{k}{\kappa}\right)^2
    \left(1-\frac{T}{T'}\right)^2.
\end{align}
The total EP rate is just the sum of them:
\begin{align}
    \sigma^\tot
    =\frac{2\kappa^2}{k}\frac{T'}{T}
    \left[1+\left(\frac{k}{\kappa}\right)^2
    \left(1-\frac{T}{T'}\right)^2\right].
\end{align}
These are plotted in \fref{fig: 2dflow} (c). {$\sigma^\ex$ and $\sigma^\tot$ diverges at $T'\rightarrow0$ because the system state takes the form of the delta-function distribution which is an extremely nonequilibrium state.}

\section{Details of the example: rocking ratchet}
In this section, we present some calculations and figures to supplement the second example in the main text.
\subsection{Model}
The Langevin equation of the rocking ratchet is
\begin{align}
    \dot{x}=F(x,t)+\xi(t)=-\del_x U(x)+R(t)+\xi(t),
\end{align}
where the potentials are periodic in space and time: $U(x+L)=U(x)$ and $R(t\calT)=R(t)$.
The thermal noise $\xi(t)$ is the white Gaussian noise, i.e., $\bkt{\xi(t)}=0$ and $\xi(t)\xi(t')=\delta(t-t')$. The corresponding Fokker-Planck equation is 
\begin{align}
    \label{rocking: Fokker-Planck equation}
    \del_t p(x,t)=-\del_x j(x,t)=-\del_x[F(x,t)-D\del_x]p(x,t),
\end{align}
and the mean local velocity is expressed as $v(x,t):=j(x,t)/p(x,t)=F-D\del_x \ln p$. Since the system is periodic in space, the probability $p$ and the probability current $j$ are punishingly small. We assume that these quantities are also periodic, $p(x+L,t)=p(x,t), j(x,t)=j(x+L,t)$, and take the summation $\hat{p}(x, t):=\sum_{n=-\infty}^\infty p(x+nL,t),\hat{j}(x, t):=\sum_{n=-\infty}^\infty j(x+nL,t)$. Then, the reduced quantities $\hat{p}$ and $\hat{j}$ also get periodic in space and have finite values, satisfying the same form of the Fokker-Planck equation as \eref{rocking: Fokker-Planck equation}. This assumption is plausible if we consider the periodic steady state or prepare the initial state periodic in space. 
{The instantaneous steady state is given by a modified canonical distribution:
\begin{align}
    p^\ssrm(x, t)=\frac{\int^{x+L}_x dye^{[V(y,t)-V(x,t)]/D}}{\int_0^L dx\int^{x+L}_x dye^{[V(y,t)-V(x,t)]/D}}, 
\end{align}
where $V(x, t):=U(x)-R(t)x$ and $\int^{x+L}_x dye^{V(y, t)/D}$ is required to satisfy the periodicity $p^\ssrm(x+L)=p^\ssrm(x)$.}
We define the reduced local mean velocity as $\hat{v}(x, t):=\hat{j}(x,t)/\hat{p}(x,t)$ which coincides with the usual definition $v$ as long as the probability has the periodicity $p(x+L,t)=p(x,t)$. These reduced quantities are denoted as $p,j,v$ in the main text. The currents are EP rates can be calculated in the usual way, but the integration range is replaced by $[0,L]$.

\begin{figure}
    \centering
    \includegraphics[width=1.0\linewidth]{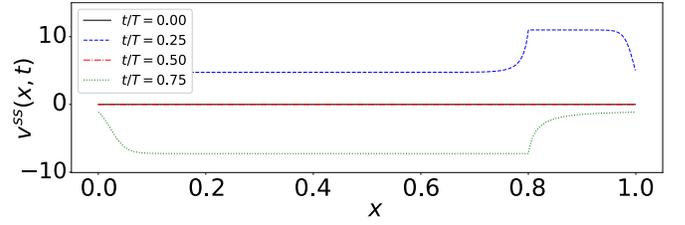}
    \caption{\label{fig: rocking_mlv} 
    The steady-state mean local velocity in the rocking ratchet model. Since the nonconservative force vanishes at $t/T=0,0.5$, the mean local velocity vanishes as well at this time. The plot has kinks at $x/L=0, 0.8$ in accordance with the indifferentiability of the potential $U(x)$. Apart from these points, the plot is approximately flat. The parameters used in calculation are $D = 0.1, T=0.7, a=0.8, R_0 = 6.0$.
    }
\end{figure}

\subsection{The coefficient choice}
In the main text, we used the particle current with $d(x)=1$. Here, we consider another choice of coefficient. We plot the mean local velocity for the instantaneous steady state in \fref{fig: rocking_mlv}. $v^\ssrm$ tends to be constant except for the top and valley of the potential $U(x)$, which suggests that the coefficient $d(x)=1$ approximately satisfies the condition $d\propto e^\hk$ (\eref{hk coeff}).
Consequently,$j_d^\ex$ is small compared to $j_d^\hk$ and the estimation of the excess EP rate does not have high accuracy. (see \fref{fig: rocking} (a) and (d)). 

\begin{figure}
    \centering
    \includegraphics[width=1.0\linewidth]{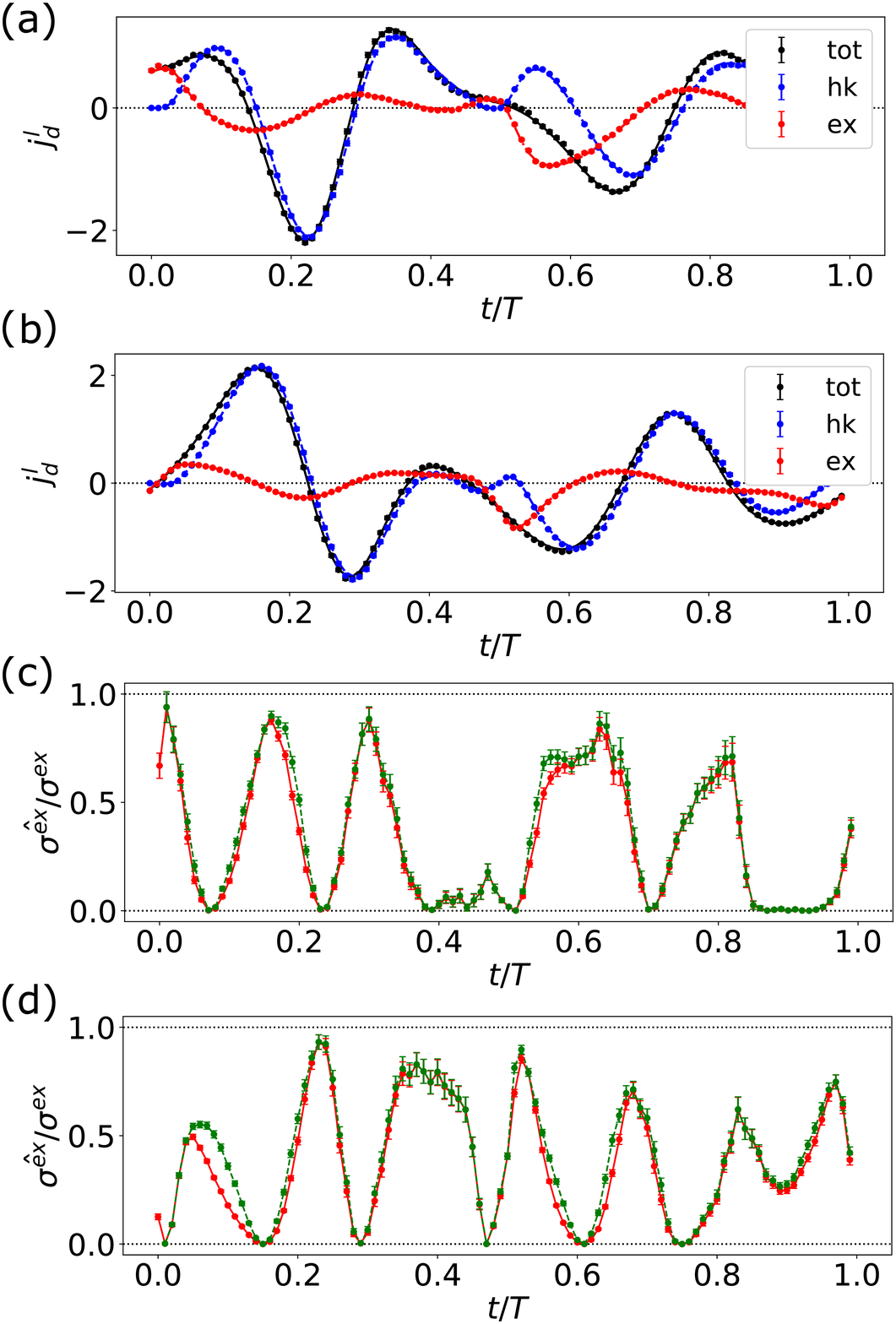}
    \caption{\label{fig: rocking_cossin} 
    (a)(b) Each current average with $d(x)=\cos(2\pi x/L)$ for (a) and $d(x)=\sin(2\pi x/L)$ for (b).  (c)(d) The estimation of the excess EP rate based on \eref{projective TUR} (green line) and \eref{respective TUR} (red line) with $d(x)=\cos(2\pi x/L)$ for (c) and $d(x)=\sin(2\pi x/L)$ for (d).
    The parameter used in calculation are $D = 0.1, T=0.7, a=0.8, R_0 = 6.0$.
    }
\end{figure}

Since the mean local velocity is expected to consist of low-frequency Fourier components, we use the coefficient $d(x)=\cos(2\pi x/L),\sin(2\pi x/L)$ instead here. We plot the current averages in \fref{fig: rocking_cossin}(a)(b). The housekeeping current is decreased in comparison with the case $d(x)=1$, and is now comparable to the excess current. We can see the improvement of the estimation in the first half of the period (see \fref{fig: rocking_cossin}(c)(d)). 
Although $\cos(2\pi x/L)$ and $\sin(2\pi x/L)$ are not orthogonal to each other in terms of the inner product $\bkt{\cdot,\cdot}$, the currents and estimations tend to compensate each other. Combining these results, we can estimate roughly over $50\%$ of the excess EP rate. 

\section{2-beads model}
In the first example in the main text, the direction of $\bse^\hk,\bse^\ex$ or $\bsv^\ssrm,\bsv-\bsv^\ssrm$ was obvious. Also, as noted in the discussion, the direction of $\bsv^\ssrm$ is considered experimentally estimable. In this section, we consider a two-dimensional heat conduction system in the absence of prior knowledge of the steady state.

\subsection{Setup}
\begin{figure}
    \centering
    \includegraphics[width=1.0\linewidth]{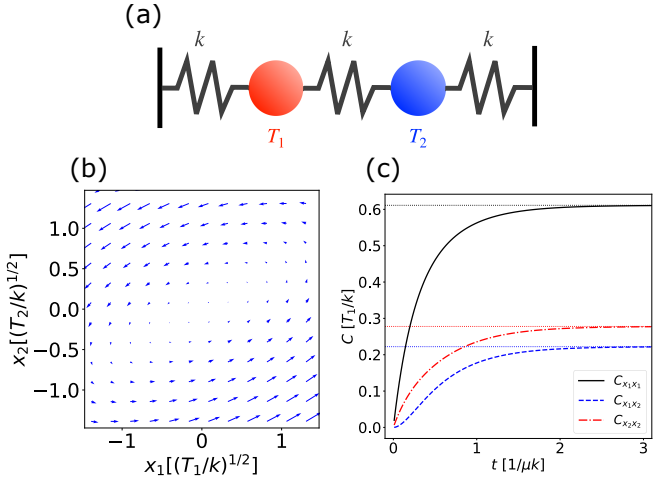}
    \caption{\label{fig: 2-beads_model}
    (a) The sketch of the 2-beads model. The two beads (particles) are coupled to each other. The left (right) bead is influenced by thermal fluctuation from the hot (cold) reservoir. (b) The steady-state mean local velocity, $\bsv^\ssrm$. $\bsv^\ssrm$ is divided by $\mu k$ to be dimensionless, and the scale is multiplied by $1/10$ for illustration. (c) The time-evolution of the covariance matrix of the state, \eref{2-beads: state distribution}. The parameter used in calculation are $\mu=1.0, k=1.0, T_1=1.5, T_2=0.5$.
    }
\end{figure}

Consider two coupled Brownian particles governed by the Langevin equation
\begin{align}
    \label{2-beads: Langevin eq}
    \dot{\bsx}&=\mu A\bsx+\sqrt{2}G\bsxi(t)\\
    &\textrm{with  }
    A=\mu k\begin{bmatrix}
                -2 & 1\\
                1 & -2
            \end{bmatrix}\ 
    \textrm{and } G=\begin{bmatrix}
                \sqrt{\mu T_1} & 0\\
                0 & \sqrt{\mu T_2}
            \end{bmatrix}
\end{align}
and the Fokker-Planck equation
\begin{align}
    \label{2-beads: Fokker-Planck eq}
    \partial_t &p(\bsx, t)=-\bsna^T\bsj(\bsx, t)\\
    \label{2-beads: probability flow}
    &\textrm{with } \bsj(\bsx, t)=(A\bsx-D\bsna)p(\bsx, t).
\end{align}
The particles undergo thermal fluctuations interacting with a heat bath at temperatures $T_1, T_2\ (T_1 > T_2)$, respectively (see \fref{fig: 2-beads_model}(a)). This model has also been investigated in the literature \cite{Li-Horowitz-Gingrich-Fakhri2019,Otsubo-Ito-Dechant-Sagawa2020} and others. Here, we obtain the distribution of states according to the literature \cite{Li-Horowitz-Gingrich-Fakhri2019}.
Provided that the steady-state distribution is expressed as the Gaussian distribution
\begin{align}
    p^\ssrm(\bsx)=\frac{1}{2\pi\sqrt{\det \bar{C}}}
    \exp \left[-\frac{1}{2}\bsx^T\bar{C}^{-1}\bsx\right]
\end{align}
with a symmetric correlation matrix $\bar{C}$, the steady-state probability distribution is given by
\begin{align}
    \bsj^\ssrm=(A+D\bar{C}^{-1})\bsx p^\ssrm.
\end{align}
At this time, $\bar{C}$ satisfies the Lyapunov equation \cite{lyapunov1992general}
\begin{align}
    \label{2-beads: Lyapunov eq}
    A\bar{C}+\bar{C}A+2D=0.
\end{align}
The steady-state mean local velocity is plotted in \fref{fig: 2-beads_model}(b). Unlike in the case of a 2-dimensional vortex system, the flow is elliptical and its shape depends on the temperature of the heat bath. Therefore, it is difficult to determine the direction of $\bsv^\ssrm$ from physical considerations. The current defined below does not satisfy the condition of \eref{hk coeff}\eqref{ex coeff} and is not reduced to the total current average, and hence it is difficult to measure them experimentally.

The solution of the Langevin equation (\eref{2-beads: Langevin eq}) is 
\begin{align}
    \bsx(t)=\int^t_0 ds e^{A(t-s)}\sqrt{2}G\bsxi(s)+e^{At}\bsx(0)\ (t\geq0).
\end{align}
Assuming $\bkt{\bsx(0)}=0$, the correlation matrix at time $t$ is calculated as
\begin{align}
    \label{2-beads: corr mat}
    C(t):&=\bkt{\bsx(t)\bsx(t)^T}\nonumber\\
        &=2\int^t_0 ds e^{A(t-s)}D^{A(t-s)}+e^{At}C(0)e^{At}.
\end{align}
This is plotted in \fref{fig: 2-beads_model}(c).
If we adopt the Gaussian distribution as the initial state, the state at time $t$ is given by the Gaussian distribution whose covariance matrix is $C(t)$:
\begin{align}
    \label{2-beads: state distribution}
    p(\bsx, t)=\frac{1}{2\pi\sqrt{\det C(t)}}
    \exp \left[-\frac{1}{2}\bsx^T{C(t)}^{-1}\bsx\right].
\end{align}
Since this is the Gaussian distribution, the average of the quantity in the shape of $\bsx^T H\bsx$ is computed as ${\bsx^T H\bsx}=\textrm{tr}(C^{1/2}HC^{1/2})=\textrm{tr}(HC)$. The second term on the right-hand side of \eref{2-beads: corr mat} vanishes in the long-time limit because the all eigenvalues of $A$ are negative. By contrast, the first term, denoted as $C_0(t)$, satisfies \eref{2-beads: Lyapunov eq} due to 
\begin{align}
    \frac{d}{dt}C_0(t)=AC_0+C_0A+2D\xrightarrow{t\rightarrow\infty}0.
\end{align}
Therefore, the correlation matrix in the steady state is 
\begin{align}
    \bar{C}=\lim_{t\rightarrow\infty}C_0(t)
    =\frac{1}{12k}
    \begin{bmatrix}
        7T_1+T_2 & 2(T_1+T_2)\\
        2(T_1+T_2) & T_1+7T_2
    \end{bmatrix}.
\end{align}

\subsection{Entropy production and current}
\begin{figure}
    \centering
    \includegraphics[width=1.0\linewidth]{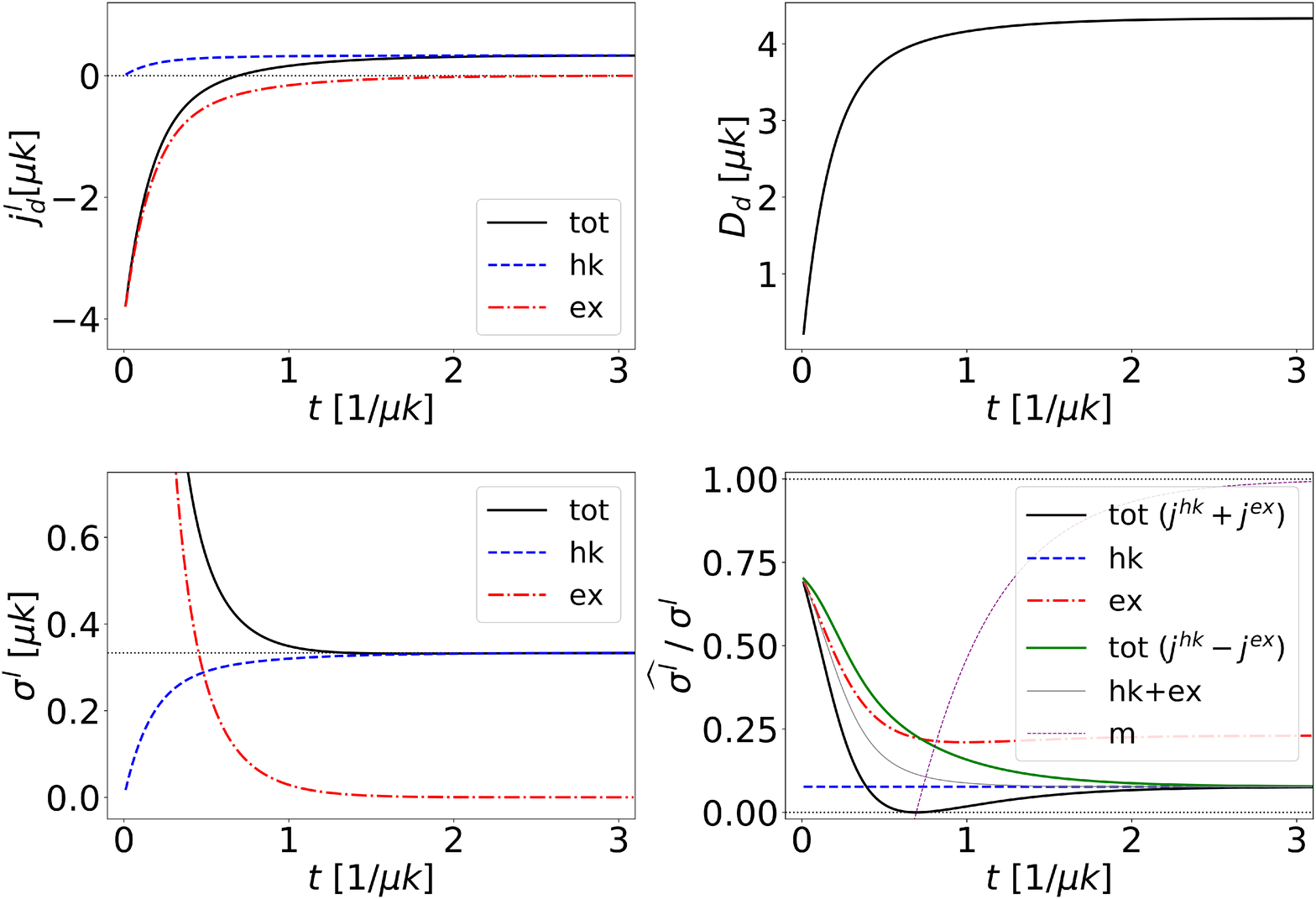}
    \caption{\label{fig: 2-beads_TUR}
    (a) The current averages in the 2-beads model. The current coefficient is $\bsd=D^{-1}A\bsx$ so that the total current agrees with the EP rate of the reservoirs. The initial state is the delta function localized at the origin. Therefore, at first, the total current average is negative because the energy is received as heat from the heat reservoir. Eventually, when it reaches the steady state, the excess current average vanishes and the housekeeping current average agrees with that of the total current. Since there is a steady heat flow from the high-temperature heat reservoir $T_1$ to the low-temperature heat reservoir $T_2$, the steady-state value of the current average is positive. (b) The current variance. This is initially zero because they are localized at the origin and increases as the particles diffuse due to thermal fluctuations, eventually reaching the steady-state value. (c) The EP rates. The total EP rate diverges at the initial time and decreases monotonically over time, saturating to a finite steady-state value. The excess EP rate is similar but has a steady-state value of $0$. On the other hand, the housekeeping EP rate monotonically increases from $0$ to reach the same steady-state value as the total system. (d) The estimation of each EP rate based on \eref{respective TUR}\eqref{max TUR}. The black and green lines correspond to the estimation $\widehat{\sigma^\tot}=(j_d^\hk+j_d^\ex)^2/\calD_d, (j_d^\hk-j_d^\ex)^2/\calD_d$ respectively. The gray line represents the sum of the estimation $\widehat{\sigma^\tot}=[(j_d^\hk)^2+(j_d^\ex)^2]/\calD_d$. The purple line shows the EP rate of the reservoirs $\widehat{\sigma^\tot}=j_d=\sigma^m$. The parameters used in calculation are $\mu=1.0, k=1.0, T_1=1.5, T_2=0.5$.
    }
\end{figure}
Now that we are ready, we calculate the currents and the entropy productions. The initial state is given by a delta function localized at the origin. This gives $C(0)=0$. Choosing $\bsd=D^{-1}A\bsx$ as the coefficient, the current is the EP rate in the heat bath. Each current average is 
\begin{align}
    j_d&=
    \textrm{tr}\left(AD^{-1}(A+DC^{-1})C\right),\\
    j_d^\hk&=
    \textrm{tr}\left(AD^{-1}(A+D\bar{C}^{-1})C\right),
\end{align}
and 
\begin{align}
    j_d^\ex=
    j_d-j_d^\hk=
    \textrm{tr}\left(A(C^{-1}-\bar{C}^{-1})C\right).
\end{align}
The current variance is computed as
\begin{align}
    \calD_d=\textrm{tr}(AD^{-1}AC).
\end{align}
These are plotted in \fref{fig: 2-beads_TUR}(a) and (b). Since the minimum energy of the system is $0$ at $t=0$, the energy flows into the system as heat, leading to $j_d<0$. For the housekeeping current, $j_d^\hk$ vanishes in the limit $t\rightarrow0$ because of $C\rightarrow0$. At $t\rightarrow\infty$, there is a steady heat flow and  $j_d^\hk\rightarrow j_d>0$. Due to this combination, the housekeeping current is always above $0$ and the excess current is always below $0$. 
In the steady state $(t\rightarrow\infty)$, $j_d^ex$ vanishes because of $C\rightarrow\bar{C}$. On the other hand, when the detailed balance condition holds $(T_1=T_2)$, $D\propto I$ leads to $\bar{C}=-DA^{-1}=-A^{-1}D$ and $j_d^\hk$ vanishes.
As for fluctuations, $\calD_d$ increases monotonically and saturates to a stationary value as time passes and particles diffuse from near the origin.

The EP rates are computed as
\begin{align}
    \label{2-beads: tot EP rate}
    {\sigma^\tot}&=
    \textrm{tr}\left((A+DC^{-1})^TD^{-1}(A+DC^{-1})C\right),\\
    \label{2-beads: hk EP rate}
    {\sigma^\hk}&=
    \textrm{tr}\left((A+D\bar{C}^{-1})^TD^{-1}(A+D\bar{C}^{-1})C\right),
\end{align}
and
\begin{align}
    \label{2-beads: ex EP rate}
    {\sigma^\ex}&=
    {\sigma^\tot}-{\sigma^\hk}\nonumber\\
    &=\textrm{tr}\left((C^{-1}-\bar{C}^{-1})^TD(C^{-1}-\bar{C}^{-1})C\right).
\end{align}
We plot them in \fref{fig: 2-beads_TUR} (c). First, the total and excess EP rates diverge in the limit $t\rightarrow0$ because they include $C^{-1}$ which does not cancel out with $C\rightarrow0$, and the housekeeping EP rate with only $C$ terms vanishes in this limit. As in the previous example (2-dimensional flow), the excess EP rate diverges for the localized nonequilibrium state of the delta function and the housekeeping EP rate becomes $0$. 
Over time, the housekeeping EP rate monotonically increases and saturates, and the excess EP rate decays to zero. 
In the steady state, since the EP rate of the system is zero, the steady-state value reached by the total and housekeeping EP rates is the same as the corresponding steady-state value of the current average (see \fref{fig: 2-beads_TUR}(a)).

Finally, the estimation of the EP rates by TUR is plotted in \fref{fig: 2-beads_TUR} (d). For the EP rate of the total system, the green line is above the black line because the product of the currents is always negative in this setting (see \fref{fig: 2-beads_TUR}(a)). Around the initial time, the accuracy of the estimation using \eref{max TUR} is relatively high. As it approaches the steady state, the accuracy becomes worse, indicating that it is better to use the current as it is, i.e., the EP rate of the heat bath (purple line) for the estimation. This is because the mean of the current asymptotically approaches the mean of the EP rate for the total system, but the variance does not, making the estimation with TUR less accurate.

Since $j_d^\hk$ vanishes in the limit $t\rightarrow0$, all but the red line (and the purple line) converge to the same point. By contrast, $j_d^\ex$ vanishes in the limit $t\rightarrow\infty$, and all but the red line (and the purple line) intersect at the single point.

Around the initial time, the estimation of the excess EP rate has a similar trend to that of the total system. The accuracy deteriorates as it approaches the steady state, but it is obvious that it relaxes to zero in this case. On the other hand, the estimation of the housekeeping EP rate tends to be similar to that of the total system when the system is close to the steady state and is not accurate in any time region.

%

\end{document}